\documentclass[preprint,floatfix] {revtex4} 
 
\usepackage{graphicx}
\usepackage{subfigure}
\begin{document}

\title{Studies on some singular potentials in quantum mechanics}
\author{Amlan K. Roy}
\affiliation{Department of Chemistry, University of New Brunswick, 
Fredericton, NB, E3B 6E2, Canada}
\email{akroy@chem.ucla.edu}
\begin{abstract}
A simple methodology is suggested for the efficient calculation of certain central 
potentials having singularities. The generalized pseudospectral method used in this work 
facilitates {\em nonuniform} and optimal spatial discretization. Applications have been 
made to calculate the energies, densities and expectation values for two singular 
potentials of physical interest, {\em viz.,} (i) the harmonic potential plus inverse 
quartic and sextic perturbation and (ii) the Coulomb potential with a linear and quadratic 
term for a broad range of parameters. The first 10 states belonging to a maximum of 
$\ell=8$ and 5 for (i) and (ii) have been computed with good accuracy and compared with the
most accurate available literature data. The calculated results are in excellent agreement,
especially in the light of the difficulties encountered in these potentials. Some new 
states are reported here for the first time. This offers a general and efficient scheme for
calculating these and other similar potentials of physical and mathematical interest in 
quantum mechanics accurately. 
\end{abstract}
\maketitle

\section{Introduction}
Application of quantum mechanics to many branches of physics and chemistry, e.g., the 
atomic and molecular physics, nuclear physics, particle physics, solid-state physics, 
astrophysics, etc., often involves a potential term having {\em singularity} and usually 
some extra perturbation terms characterizing the physical system under study. Almost for 
all practical purposes, the radial Schr\"odinger equation to be solved requires 
approximations as the exact solutions can be obtained only for a few idealized cases such 
as the harmonic oscillator or the Coulomb potential. Consequently, a large number of 
analytical and numerical methodologies have been developed by many workers over the decades 
to obtain accurate solutions for such systems employing a variety of techniques. Although 
many attractive and elegant approaches exist in the literature for both the non-singular 
and singular potentials, the latter remains relatively less explored as one encounters 
considerable difficulties and challenges because of the singularity. Therefore, the search 
for a {\em general} scheme capable of producing accurate and reliable results, has 
continued to remain an active area of research.  

The purpose of this article is to assess the performance of a simple numerical methodology 
for the singular potentials. Although the method would hold good for other singularities, 
we will restrict ourselves to two different classes of such potentials which have found
applications in various branches of physics, {\em viz.,}\\ 
(a) the harmonic potential including an inverse quartic and sextic 
anharmonicity,
\begin{equation}\label{eqn:pot1}
V(r)=a_1 r^2+\sum_{i=2}^{3} a_i r^{-2i}, \ \ \ \  (a_1, a_3 > 0)
\end{equation}
and \\
(b) the Coulomb potential with a linear and quadratic coupling,
\begin{equation}\label{eqn:pot2}
V(r)=-b_1/r+\sum_{j=1}^{2} b_j r^j 
\end{equation}
In what follows we will adopt $a_1=a,$ $a_2=b,$ $a_3=c;$ and $b_1=Z,$ $b_2=g,$ 
$b_3=\lambda$ for the sake of consistency with the literature. 

The potential in Eq.~(\ref{eqn:pot1}) is defined in $r \in [0, \infty]$ with Dirichlet 
boundary conditions. Such a potential may lead to many interesting and appealing features; 
both physically and mathematically. For example, there may be the so-called \emph {Klauder 
phenomenon}[1-2] signifying that once the singular perturbation is turned on, it cannot be 
completely turned off. It is noticed that the effects of perturbation is never {\em small} 
enough to ignore. For example, the complete three-dimensional Schr\"odinger equation is 
exactly solvable without the inverse sextic perturbation in Eq.~(\ref{eqn:pot1}) for both 
the extreme cases $a=0$ and $b=0.$ However subjecting the system to an infinitesimal 
perturbation $a \ne 0, b \ne 0$ induces such dramatic effects in the wave function near the 
origin that the perturbation theory fails to give {\em any} estimate at all [3]. On the 
other hand, when either $b$ or $c$ is large, the anharmonic terms compete with the harmonic 
term and none of the interacting terms can be considered negligible. Several methods are 
available for the calculation of the eigenvalues and eigenfunctions of this potential. Some
of them, for example, include the {\em singular} perturbation theory [4-6] specially 
designed for this potential, variational methods [7,8], step-by-step analytic continuation 
method [9], numerical solution using finite-difference scheme [10,11] or the B spline 
techniques [12], etc. This potential also has the importance in the context of conditional 
exact solvability; in other words, exact analytic solutions can be obtained for some 
severely restricted set of parameters (related to each other through some special relations) 
for ground states [13] or excited states [14,15]. 

The potential in Eq.~(\ref{eqn:pot2}) also has many relevance to physics. Thus, the Coulomb 
plus the linear potential has been studied extensively for the spherical Stark effects in 
hydrogenic atoms [16], or in connection with the various non-relativistic quark confinement 
and other similar problems in particle physics [16-18], etc. On the other hand, the Coulomb 
plus the quadratic potential has been studied in the context of quadratic Zeeman effects in 
hydrogenic atoms [19], or in plasma physics [20], etc. Again various methodologies have 
been devised for the calculation of eigenvalues of this potential by many workers. For some 
special values of the parameters $Z, g$ and $\lambda,$ this leads to a quasi exactly 
solvable potential and analytic solutions can be obtained [21,22]. Other works include 
Rayleigh-Schr\"odinger perturbation model [23], the Stieltjes moments method [21], the 
analytic continued fraction theory [24,25], the Hill-determinant method [26], the 
supersymmetric quantum mechanics coupled with the shifted 1/N method [22,27], the two-point 
quasifractional approximant method [28], etc. More references on this potential can be 
found in [28]. 

However, for both these potentials, good accuracy results can be achieved only by a few of 
the above mentioned methods. Thus leaving aside a few methods, (like [9] or [11], for 
example) it is usually quite difficult to reach beyond a six- or seven-decimal place 
accuracy for the potential in Eq.~(\ref{eqn:pot1}) for an arbitrary set of parameters. As 
another author [8] points out, to obtain even this much accuracy by standard numerical 
methods for the potential like that in Eq.~(\ref{eqn:pot1}) sometimes requires one to use a 
mesh of at least 80,000 points for some cases. Also, some of these methods, although can 
provide good quality results for certain type of parameters, perform rather poorly or even 
can not provide any result at all in other occasions (see [22], for example). Finally while 
majority of the works have remained largely focused to the ground states, relatively less 
attention has been paid to the excited states (especially those associated with the higher 
angular momentum). Given these facts, a general method which can offer accurate as well as 
reliable results for a general set of parameters for both ground and excited states (low as 
well as high) with equal ease, would be highly desirable and demanding. This work attempts 
to make a small step in such a direction. The seed of the motivation grew from a study of 
singly- and doubly- and triply excited Rydberg states of many-electron atomic systems 
[29,30] within the density functional framework [31] using the generalized pseudospectral 
(GPS) [32,33] scheme which produced good quality results within the bounds of the theory. 

The purpose of this article is to use this simple numerical scheme for the calculation of 
the above mentioned singular potentials. First we will present in brief the essentials of 
the GPS method for the solution of the single-particle radial Schr\"odinger equation in 
section II. This has witnessed many successful applications in electronic structure and 
dynamics calculations in the recent years involving mainly Coulomb singularities (see, for 
example, [29-30, 32-35]). Thereafter this has been extended to some other physical systems
including the spiked harmonic oscillators, the logarithmic and power potentials as well as
the Hulthen and Yukawa potentials [36-38] with considerable promise. Section III presents 
the computed eigenvalues, wave functions, radial densities and the expectation values for 
the tow cases. Both ground as well as higher excited states are calculated and a thorough 
comparison with literature data has been made, wherever possible. Finally few conclusions 
are drawn in section IV.

\section{Methodology}\label{sec:method}
In this section, we present an overview of the generalized pseudospectral method (GPS) 
employed to solve the radial eigenvalue problem with the singular potentials. A more 
detailed account can be found in the refs. [29-30, 32-38].

Without loss of generality, the desired radial Schr\"odinger equation can be written as 
(atomic units employed unless otherwise mentioned),
\begin{equation}
\hat{H}(r)\ \phi(r) =\varepsilon \ \phi(r),
\end{equation}
where the Hamiltonian includes the usual kinetic and potential energy operators,  
\begin{equation}
\hat{H}(r) =-\frac{1}{2} \ \ \frac{d^2}{dr^2} +v(r),
\end{equation}
with 
\begin{equation}
 v(r) = V(r) + \frac{\ell (\ell+1)}{2r^2} 
\end{equation}
and $V(r)$ is given by Eq.~(\ref{eqn:pot1}) or (\ref{eqn:pot2}). The symbols have their 
usual significances. The usual finite-difference spatial discretization schemes often 
require a large number of grid points to achieve good accuracy since majority of these 
methods employ a uniform mesh (nonuniform schemes are used in a few occasions as well, 
e.g., in [11]). The GPS method, however, can give \emph {nonuniform} and optimal spatial 
discretization accurately. This allows one to work with a denser mesh at shorter $r$ 
regions and a coarser mesh at larger $r$. Additionally the GPS method is computationally 
orders of magnitude faster than the finite-difference schemes.  

One of the principal features of this scheme lies in the fact that a function $f(x)$ 
defined in the interval $x \in [-1,1]$ can be approximated by the polynomial $f_N(x)$ of 
order N so that,  
\begin{equation}
f(x) \cong f_N(x) = \sum_{j=0}^{N} f(x_j)\ g_j(x),
\end{equation}
and the approximation is \emph {exact} at the \emph {collocation points} $x_j$, i.e.,
\begin{equation}
f_N(x_j) = f(x_j).
\end{equation}
In this work, we have employed the Legendre pseudospectral method using $x_0=-1$, $x_N=1$, 
where $x_j (j=1,\ldots,N-1)$ are obtainable from the roots of the first derivative of the 
Legendre polynomial $P_N(x)$ with respect to $x$, i.e., 
\begin{equation}
P'_N(x_j) = 0.
\end{equation}
The cardinal functions, $g_j(x)$ in Eq.~(6) are given by the following expression,
\begin{equation}
g_j(x) = -\frac{1}{N(N+1)P_N(x_j)}\ \  \frac{(1-x^2)\ P'_N(x)}{x-x_j},
\end{equation}
obeying the unique property $g_j(x_{j'}) = \delta_{j'j}$. Now the semi-infinite domain 
$r \in [0, \infty]$ is mapped into the finite domain $x \in [-1,1]$ by the transformation 
$r=r(x)$. One can make use of the following algebraic nonlinear mapping,
\begin{equation}
r=r(x)=L\ \ \frac{1+x}{1-x+\alpha},
\end{equation}
where L and $\alpha=2L/r_{max}$ may be termed as the mapping parameters. Now, introducing 
the following relation, 
\begin{equation}
\psi(r(x))=\sqrt{r'(x)} f(x)
\end{equation}
coupled with the symmetrization procedure [32,33] leads to the transformed Hamiltonian as 
below, 
\begin{equation}
\hat{H}(x)= -\frac{1}{2} \ \frac{1}{r'(x)}\ \frac{d^2}{dx^2} \ \frac{1}{r'(x)}
+ v(r(x))+v_m(x),
\end{equation}
where $v_m(x)$ is given by,
\begin{equation}
v_m(x)=\frac {3(r'')^2-2r'''r'}{8(r')^4}.
\end{equation}
Note the advantage that this leads to a \emph {symmetric} matrix eigenvalue problem which 
can be readily solved to give accurate eigenvalues and eigenfunctions. For the particular 
transformation used in Eq.~(10), $v_m(x)=0$. This discretization then leads to the 
following set of coupled equations, 
\begin{widetext}
\begin{equation}
\sum_{j=0}^N \left[ -\frac{1}{2} D^{(2)}_{j'j} + \delta_{j'j} \ v(r(x_j))
+\delta_{j'j}\ v_m(r(x_j))\right] A_j = EA_{j'},\ \ \ \ j=1,\ldots,N-1,
\end{equation}
\end{widetext}
where
\begin{equation}
A_j  = \left[ r'(x_j)\right]^{1/2} \psi(r(x_j))\ \left[ P_N(x_j)\right]^{-1}.
\end{equation}
and the symmetrized second derivative of the cardinal function, $D^{(2)}_{j'j}$ is given by,
\begin{equation}
D^{(2)}_{j'j} =  \left[r'(x_{j'}) \right]^{-1} d^{(2)}_{j'j} 
\left[r'(x_j)\right]^{-1}, 
\end{equation}
with
\begin{eqnarray}
d^{(2)}_{j',j} & = & \frac{1}{r'(x)} \ \frac{(N+1)(N+2)} {6(1-x_j)^2} \ 
\frac{1}{r'(x)}, \ \ \ j=j', \nonumber \\
 & & \nonumber \\
& = & \frac{1}{r'(x_{j'})} \ \ \frac{1}{(x_j-x_{j'})^2} \ \frac{1}{r'(x_j)}, 
\ \ \ j\neq j'.
\end{eqnarray}
The performance of the method has been tested for a large number of potentials with many 
other works in the literature with respect to the variation of the mapping parameters. The 
results have been reported only up to the precision that maintained stability with respect 
to these variations. Thus for all the calculations done in this work, a consistent set for 
the numerical parameters ($r_{max}=200,$ $\alpha=25$ and $N=300$) has been used which seemed 
to be appropriate for the current problem. 

\section{Results and discussion}
In order to show the efficacy of the method, we first present some specimen results for a 
few odd- and even-parity high excited states of the pure three-dimensional quartic 
oscillator corresponding to the large vibrational quantum numbers $\mathit{v}=48, 49$ and 
the angular momentum quantum numbers $\ell=0,1,\cdots ,9$ (these two quantum numbers must 
have the same parity). This was chosen because it has been quite extensively investigated by
many authors over a long period and fairly accurate results are available making it amenable 
to easy comparison. At this point, it may be noted that all the results reported in this and
all other subsequent tables throughout this article, have been {\em truncated} and not {\em 
rounded-off}. Therefore all the entries in the tables are to be taken as correct up to the 
place they are reported. The results used for comparison in Table~\ref{tab:table1} are 
chronologically: (a) the linear variation method involving diagonalization of matrices of 
large order ($800 \times 800$) [39], (b) the analytical formula based on the scaled 
oscillator approach [40], (c) the finite-difference calculation [10] and (d) the asymptotic 
shooting method [41]. As seen, the present result for all these states match exactly up to 
the 9th decimal place with the accurate results of [41], demonstrating that higher excited 
states can be obtained fairly accurately and reliably. 

\begingroup
\squeezetable
\begin{table} 
\caption {\label{tab:table1}Calculated energies (in a.u.) of the three-dimensional quartic 
oscillator for some high-lying states corresponding to $\ell=0,1,2$ and $\mathit{v}=48,49.$}
\begin{ruledtabular}
\begin{tabular} {cclllll}
{\em v} & $\ell$ & This work & Ref. [39] & Ref. [40] & Ref. [10] & Ref. [41]\\ 
\hline
48 & 0 & 250.183358697 & 250.183351 & 250.183369 & 250.183359  & 250.1833586971 \\
49 & 1 & 256.916238928 & 256.916220 & 256.916238 & 256.916239  & 256.9162389286 \\
48 & 2 & 250.096690608 & 250.096679 & 250.096671 & 250.096691  & 250.0966906080 \\
49 & 3 & 256.773728914 & 256.773732 & 256.77369  &             & 256.7737289146 \\
48 & 4 & 249.894552064 & 249.894545 & 249.894505 &             & 249.8945520647 \\
49 & 5 & 256.517359165 & 256.517338 & 256.517316 &             & 256.5173591656 \\
48 & 6 & 249.577151099 & 249.577138 & 249.577179 &             & 249.5771510991 \\
49 & 7 & 256.147382583 & 256.147373 & 256.14750  &             & 256.1473825836 \\
48 & 8 & 249.144812457 & 249.144801 & 249.1452   &             & 249.1448124575 \\
49 & 9 & 255.664161642 & 255.664146 & 255.66480  &             & 255.6641616427 \\
\end{tabular}
\end{ruledtabular}
\end{table}
\endgroup

\begingroup
\squeezetable
\begin{table}
\caption {\label{tab:table2}Comparison of the energies (in a.u.) with literature data for 
the singular potential in Eq.~(\ref{eqn:pot1}) with $a=0.5$. First five eigenvalues are 
presented for $\ell=0$ and 1.}
\begin{ruledtabular}
\begin{tabular}{ccccc}
   $b$  & $c$    &   $\ell=0$   &  $\ell=1$       \\   \hline 
$-5.625$ & 1.7578125 &$-0.999999999999$($-1.0$\footnotemark[1],$-1.0$\footnotemark[2]) 
                     &  0.09972656243  \\
         &           & 2.99999999999(3.0\footnotemark[1],3.0\footnotemark[2])  
                     & 3.45352640920   \\     
         &           &5.48535332842(5.485353\footnotemark[2])
                     &5.84343077651    \\
         &           & 7.79200908589  & 8.10379870369  \\
         &           & 10.0249058750  & 10.3072928767  \\
$-3.5$   & 24.5      & 3.50000000000(3.5\footnotemark[3]) & 3.75153315114 \\
         &           & 5.99788108291  & 6.20248548924  \\
         &           & 8.35808537819  & 8.53779618512  \\
         &           & 10.6486411431  & 10.8121723015  \\
         &           & 12.8960139536  & 13.0478732292  \\
 0.02041 & 0.09      & 2.04810689953(2.0481069\footnotemark[4],2.0481069\footnotemark[5])  
                     & 2.63680868564   \\
         &           & 4.24927125613(4.24927125\footnotemark[4],6.048105\footnotemark[5])  
                     & 4.73981650048   \\
         &           & 6.39227593858(6.3922759\footnotemark[4])  
                     & 6.82842134800   \\
         &           & 8.50708702884(8.507087\footnotemark[4])  
                     & 8.90686191659   \\
         &           & 10.6046536820  & 10.9776953418  \\
 0.5     & 0.5       & 2.50000000000(2.5\footnotemark[3]) & 2.93583462256  \\
         &           & 4.76648152281  & 5.12713779958  \\
         &           & 6.95840432469  & 7.27829362569  \\
         &           & 9.11294921787  & 9.40592929257  \\
         &           & 11.2443693329  & 11.5177212332  \\
 22.5    & 112.5     & 5.49999999999(5.5\footnotemark[3]) & 5.65277606191 \\
         &           & 7.97997008055  & 8.11418055620 & \\
         &           & 10.3661382958  & 10.4882632705 & \\
         &           & 12.6938328534  & 12.8072549804 & \\
         &           & 14.9811074983  & 15.0878538581 & \\  
\end{tabular}
\end{ruledtabular}
\footnotetext[1]{Exact value, ref. [15].} 
\footnotetext[2]{Ref. [42].} 
\footnotetext[3]{Exact value, ref. [7].}  
\footnotetext[4]{Ref. [12].}
\footnotetext[5]{Ref. [14].}
\end{table}
\endgroup

Now we turn our focus on to the singular potentials. Table~\ref{tab:table2} shows the first 
5 eigenvalues belonging to $\ell=0$ and 1 as obtained by the GPS method for a selected set 
of parameters for the even-power inverse anharmonic potentials. Besides the ground states, 
some of the excited states corresponding to $\ell=0$ have been reported in the literature 
and we quote them appropriately wherever possible. Exact analytical results are obtainable 
for parameters following certain relations among them for (i) ground [7,15] and (ii) excited 
states [15] (the exactly solvable conditionality, e.g., $a=0.5, b=-5.625, c=1.7578125$) and 
our results match very well with these. Both positive and negative values of the parameter 
$b$ have been considered, while the parameter $a$ has been fixed at 0.5 in all the cases. No 
results could be found for non-zero angular momentum states. For the first set, the 
numerical integration results are available for the first three states of $\ell=0$ [42], 
while exact analytical result [15] exist for the first two states of $\ell=0$. For the third 
set, there were some controversy in the literature regarding the position of the second 
state corresponding to $\ell=0$. It was estimated to be at 6.048105 in [14]; later a 
numerical B-spline basis set calculation [12] re-estimated it at 4.24927125 a.u. The current 
scheme computes this with a higher accuracy (at 4.24927125613) than before and is more in 
keeping with the latter result. For the last two sets, no results are available except the 
exact analytical values for the ground states. A wide range of parameters has been used in 
this table, and these results may be useful to check the performances of other methods.  

After comparing the GPS results for the harmonic potential with inverse even-power 
anharmonicities for ground and higher excited states, we now in Table~\ref{tab:table3} 
present the first 10 eigenvalues for the same belonging to $\ell=0,2,4,6,8$ with the 
parameter set $a=0.5,$ $b=0.5\,$ and $c=0.4.$ The reason for the choice of this particular 
set lies in the fact that this is the only set for which we could find some nonzero angular 
momentum states with good accuracy [9]. Reference [9] employed the analytic continuation 
method and those are quoted. It is evident that for the available states, our results are 
in excellent agreement with those of [9]. However, these results are available only for the 
first four states of $\ell=0,1,2,3$. These results illustrate an advantage of the present 
method to treat the ground and excited states at the same footing without any special 
consideration for excited states as often required by some of the methods in the literature. 

\begingroup
\squeezetable
\begin{table}
\caption{\label{tab:table3}The first 10 eigenvalues (in a.u.) for $\ell=0,2,4,6,8$ for the 
singular potential in Eq.~(\ref{eqn:pot1}). The parameters are: $a=0.5,$ $b=0.5\,$ and 
$c=0.4.$ Numbers in the parentheses denote the values taken from Ref. [9].}
\begin{ruledtabular}
\begin{tabular}{ccccc}
 $\ell=0$       &  $\ell=2$       & $\ell=4$       & $\ell=6$       & $\ell=8$      \\ \hline
2.46735982710   &  3.66898315916  & 5.54021470933  &  7.51634080120 & 9.50878342955 \\ 
(2.46735982710) & (3.66898315916) &                &                &               \\ 
4.72473466150   &  5.76433139697  & 7.55977095004  &  9.52189664743 & 11.5110218564 \\
(4.72473466150) & (5.76433139697) &                &                &               \\
6.91000701257   &  7.85154984749  & 9.58076033037  & 11.5278374936  & 13.5133680513 \\
(6.91000701257) & (7.85154984749) &                &                &               \\
9.05914846383   &  9.93195050802  & 11.6028526028  & 13.5341458494  & 15.5158206760 \\
(9.05914846383) & (9.93195050802) &                &                &               \\
11.1859453067   & 12.0066617412   & 13.6257807401  & 15.5408038979  & 17.5183783293 \\
13.2973083828   & 14.0765778031   & 15.6493304068  & 17.5477937216  & 19.5210395512 \\
15.3972525569   & 16.1424011188   & 17.6733301630  & 19.5550974910  & 21.5238028285 \\
17.4883415197   & 18.2046882685   & 19.6976430683  & 21.5626976193  & 23.5266665991 \\
19.5723241622   & 20.2638861627   & 21.7221597681  & 23.5705768873  & 25.5296292578 \\
21.6504533563   & 22.3203586599   & 23.7467929198  & 25.5787185402  & 27.5326891600 \\
\end{tabular}
\end{ruledtabular}
\end{table}
\endgroup

\begingroup
\squeezetable
\begin{table}
\caption {\label{tab:table4} Comparison of the calculated energies (in a.u.) with literature 
data for the singular potential in Eq.~(\ref{eqn:pot2}). First four eigenvalues are 
presented for $\ell=0$ and 1.}
\begin{ruledtabular}
\begin{tabular}{ccccc}
  $Z$  &  $g$  &$\lambda$&  $\ell=0$   &  $\ell=1$             \\  \hline
 1   & 0     &  0.1    &$-0.29608776768$,$-0.29608$\footnotemark[1] 
                       &  0.57456732342                        \\
     &       &         &  0.87913607777  & 1.5383941205        \\
     &       &         &  1.8709768364   & 2.4746029046        \\
     &       &         &  2.8225931925   & 3.3984488837        \\
 1   & 0     & 10      &  4.1501236516,4.150123\footnotemark[2] 
                       &  9.5524662112                         \\
     &       &         & 13.602643792    & 18.672142320        \\
     &       &         & 22.793852381    & 27.731460369        \\
     &       &         & 31.896276164    & 36.760042590        \\
 1   & 0     & 1000    &  59.375469050,59.37546\footnotemark[3]  
                       & 106.73670248                          \\
     &       &         & 150.17477151    & 196.69803480        \\
     &       &         & 240.33685382    & 286.48903994        \\
     &       &         & 330.25417894    & 376.19005553        \\
 8   & 1     & $1/32$  & $-31.811410973$ 
             & $-7.3750000000$,$-7.375$\footnotemark[4],$-7.375000$\footnotemark[5] \\
     &       &         & $-7.2458571746$ & $-2.0483079877$,$-2.048308$\footnotemark[5]  \\
     &       &         & $-1.9150955585$ & 0.56785978768,0.567860\footnotemark[5]  \\
     &       &         & 0.70500521291   & 2.4020644372,2.402064\footnotemark[5] \\
 10  & 5     & 1       &$-49.224345286$  &$-9.8803468590$   \\
     &       &         &$-9.2898790404$  &  0.75901704864   \\  
     &       &         &  1.4166337230   &  7.4188810789    \\   
     &       &         &  8.1281375830   & 12.855211501     \\  
\end{tabular}
\end{ruledtabular}
\footnotetext[1]{Lower and upper bounds to the eigenvalue are $-0.296088$ and $-0.296087,$ 
from ref. [21].}
\footnotetext[2]{Lower and upper bounds to the eigenvalue are 4.1501236 and 4.1501239, 
from ref. [21].}
\footnotetext[3]{Lower and upper bounds to the eigenvalue are 59.3754689 and 59.3754694, 
from ref. [21].} 
\footnotetext[4]{Exact supersymmetric result, as quoted in ref. [26].}
\footnotetext[5]{Hill-determinant method, ref. [26].}
\end{table}  
\endgroup

Now results are presented for the perturbed Coulomb potential. The basic strategy of 
presentation remains the same as earlier. First, a few low-lying states along with the 
literature results for a fairly large range of parameter sets and then higher states for a 
particular case. Table~\ref{tab:table4} shows the first four computed eigenvalues for 
$\ell=0$ and 1, of the five such sets along with other results. In this case the literature 
data seems relatively scarce and scanty. Both large and small $Z$ as well as $\lambda$ 
regions have been investigated. The first three sets keep the $Z$ and $g$ fixed at 1 and 0 
respectively varying $\lambda$ from 0.1 to 1000. There are no direct results for these to 
compare; however the lower and upper bounds for the ground-state energies are available 
[21] from the Stieltje's moments method and these are mentioned appropriately. It is 
gratifying that our results fit very nicely within the small range of the bounds. For 
$Z=8, g=1, \lambda=1/32$, the exact supersymmetric result for the lowest state of $\ell=1$ 
is $-7.375$ (as quoted in [26]). The first four states of $\ell=1$ are compared with the 
Hill-determinant results [26]. Table~\ref{tab:table5}, now gives the results for first 10 
eigenvalues corresponding to $\ell=0,2,5$ for the parameter set, $Z=12, g=1, \lambda=1/32$ 
A few results are available for $\ell=2$ including the exact supersymmetric result [26] as 
well as the Hill-determinant result [26]. The agreement in our results is seen to be 
excellent.  

\begingroup
\squeezetable
\begin{table}
\caption {\label{tab:table5}The first 10 eigenvalues (in a.u.) for $\ell=0,2$ and 5 of the 
singular potential in Eq.~(\ref{eqn:pot2}) for the three sets of parameters 
$Z=12, g=1, \lambda=1/32$.}  
\begin{ruledtabular}
\begin{tabular}{ccc}
 $\ell=0$       &        $\ell=2$                              & $\ell=5$      \\  \hline 
$-71.874422806$ & $-7.1250000000$,$-7.125$\footnotemark[1],$-7.125000$\footnotemark[2]
                                                               & 0.78955622011 \\
$-17.494216229$ & $-2.7950831896$,$-2.795083$\footnotemark[2]  & 2.3644580048  \\
$-6.8642582202$ & $-0.21588294422$,$-0.215883$\footnotemark[2] & 3.7444823185  \\
$-2.5279008987$ & 1.6823565181,1.682357\footnotemark[2]        & 4.9984881647  \\
 0.05747113632  & 3.2467031143                                 & 6.1643195425  \\
 1.9612173543   & 4.6183716246                                 & 7.2648275518  \\
 3.5304071400   & 5.8655676414                                 & 8.3148289025  \\
 4.9063624018   & 7.0257108497                                 & 9.3244751785  \\
 6.1573955416   & 8.1213708656                                 & 10.301031550  \\
 7.3210117750   & 9.1671716851                                 & 11.249883522  \\
\hline
\end{tabular}
\end{ruledtabular}
\footnotetext[1]{Exact supersymmetric result, as quoted in ref. [26].}
\footnotetext[2]{Hill-determinant result, ref. [26].}
\end{table}
\endgroup

Additionally we now give the expectation values $\langle r^{-1}\rangle$ and 
$\langle r \rangle $ for both the potentials under consideration in Table~\ref{tab:table6}. 
Three states belonging to $\ell=0$ and 1 are reported for both the potentials (one set for
each of them). While no results could be found for the former case, a few results are 
available [43] for the perturbed Coulomb potential with $Z = g = \lambda = 0.5$ and good 
accuracy is observed for these. Finally, Fig.~(1) depicts the radial probability distribution 
functions $|rR_{n\ell}|^2$ for the singular potentials in Eqs. (1) and (2) (in the left and 
right panel) for the first four states belonging to $\ell=0,1,2$ respectively. The existence 
of the required number of nodes is clearly manifest. 

\begingroup
\squeezetable
\begin{table}
\caption {\label{tab:table6}Calculated expectation values (in a.u.) along with literature 
data for comparison for the two singular potentials in Eqs.~(\ref{eqn:pot1}) and 
(\ref{eqn:pot2}) for the first three states corresponding to $\ell=0$ and 1. The numbers in 
the parentheses denote the reference values from [43].} 
\begin{ruledtabular}
\begin{tabular}{ccccll}
 $a$ &  $b$    & $c$       & $\ell$&$\langle r^{-1}\rangle$  & $\langle r \rangle $  \\ \hline
 0.5 & $-5.625$& 1.7578125 &  0    & 1.037245259   & 1.040404541    \\
     &         &           &       & 0.6106362329  & 1.883380179    \\     
     &         &           &       & 0.5127801051  & 2.390745126    \\
     &         &           &  1    & 0.9798264559  & 1.109065417    \\
     &         &           &       & 0.5975951150  & 1.946934875    \\
     &         &           &       & 0.5093809276  & 2.427714809    \\ \hline
$Z$  & $g$     &$\lambda$  &$\ell$ &               &                \\ \hline
 0.5 & 0.5     & 0.5       &  0    & 1.426727774(1.4267278)   & 0.9267277745(0.92672779) \\
     &         &           &       & 1.054078325(1.0540783)   & 1.549650851(1.5496509)   \\
     &         &           &       & 0.9037976872             & 1.990327571              \\
     &         &           &  1    & 0.8497088122(0.84970883) & 1.345281337(1.3452814)   \\
     &         &           &       & 0.7355708974(0.73557090) & 1.822100782(1.8221008)   \\
     &         &           &       & 0.6671751957             & 2.204434707              \\
\end{tabular}
\end{ruledtabular}
\end{table}
\endgroup

\begin{figure}
\begin{minipage}[c]{0.40\textwidth}\centering
\includegraphics[scale=0.25]{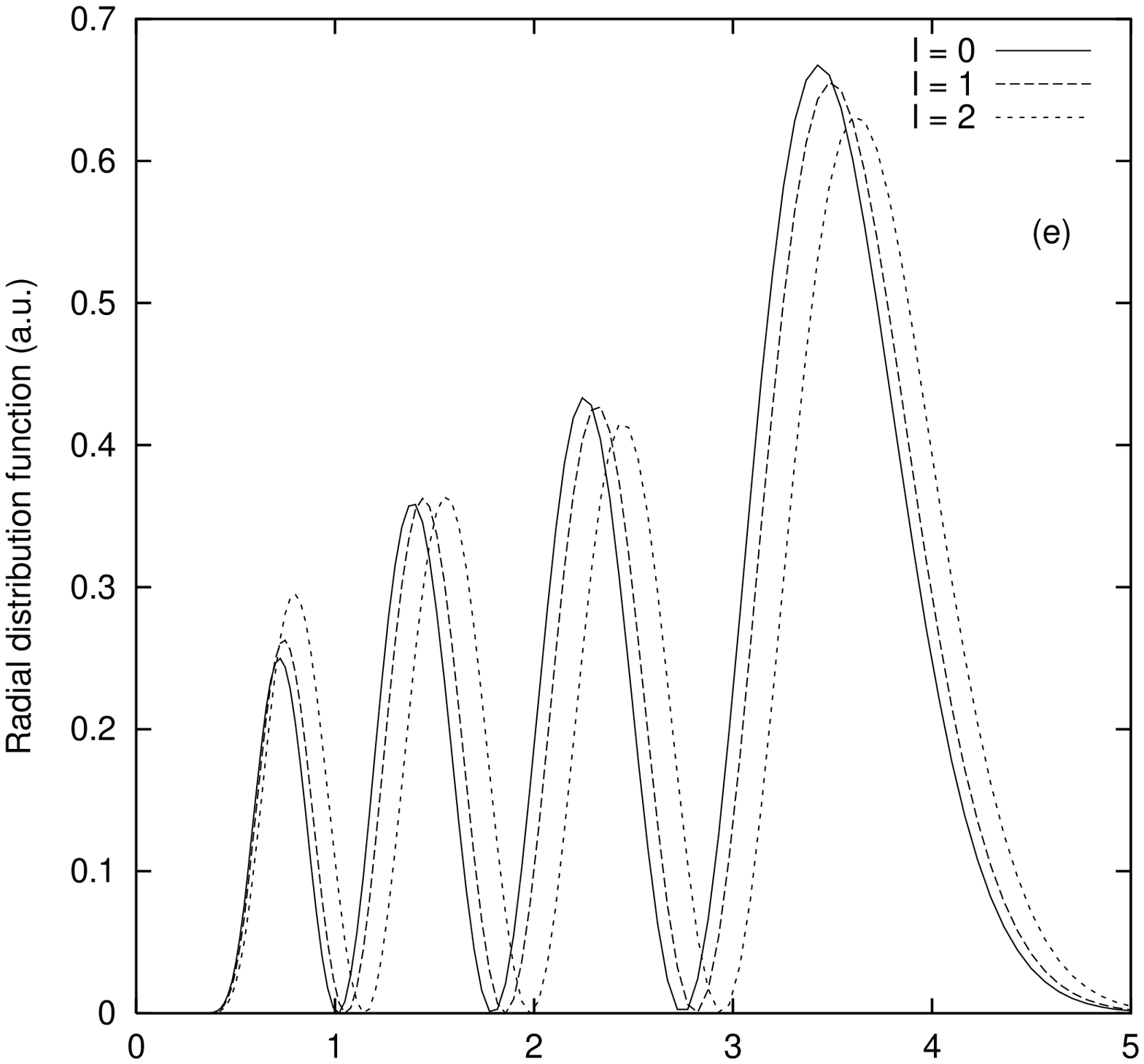}
\end{minipage}
\hspace{0.15in}
\begin{minipage}[c]{0.40\textwidth}\centering
\includegraphics[scale=0.25]{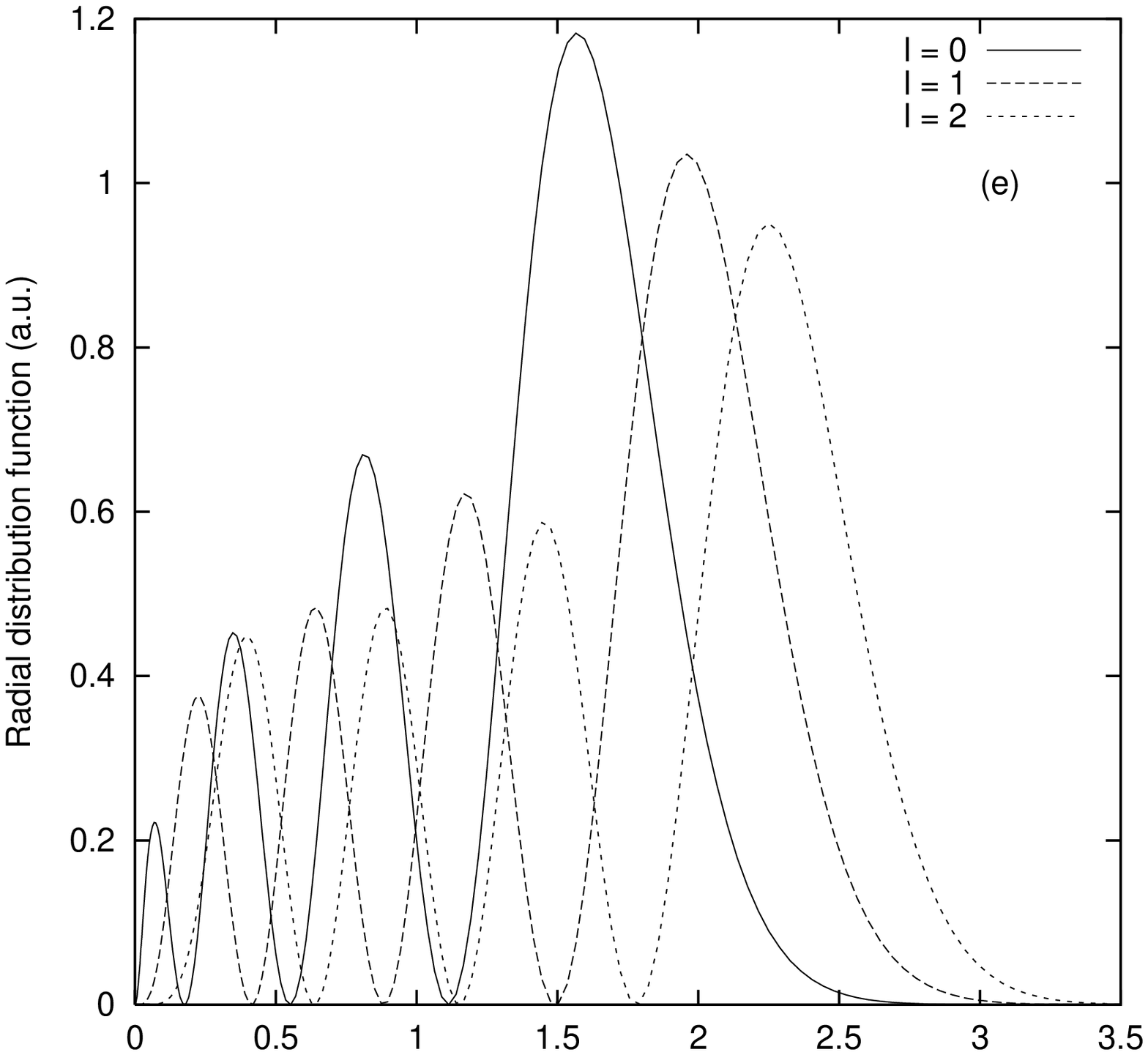}
\end{minipage}
\\
\begin{minipage}[b]{0.40\textwidth}\centering
\includegraphics[scale=0.25]{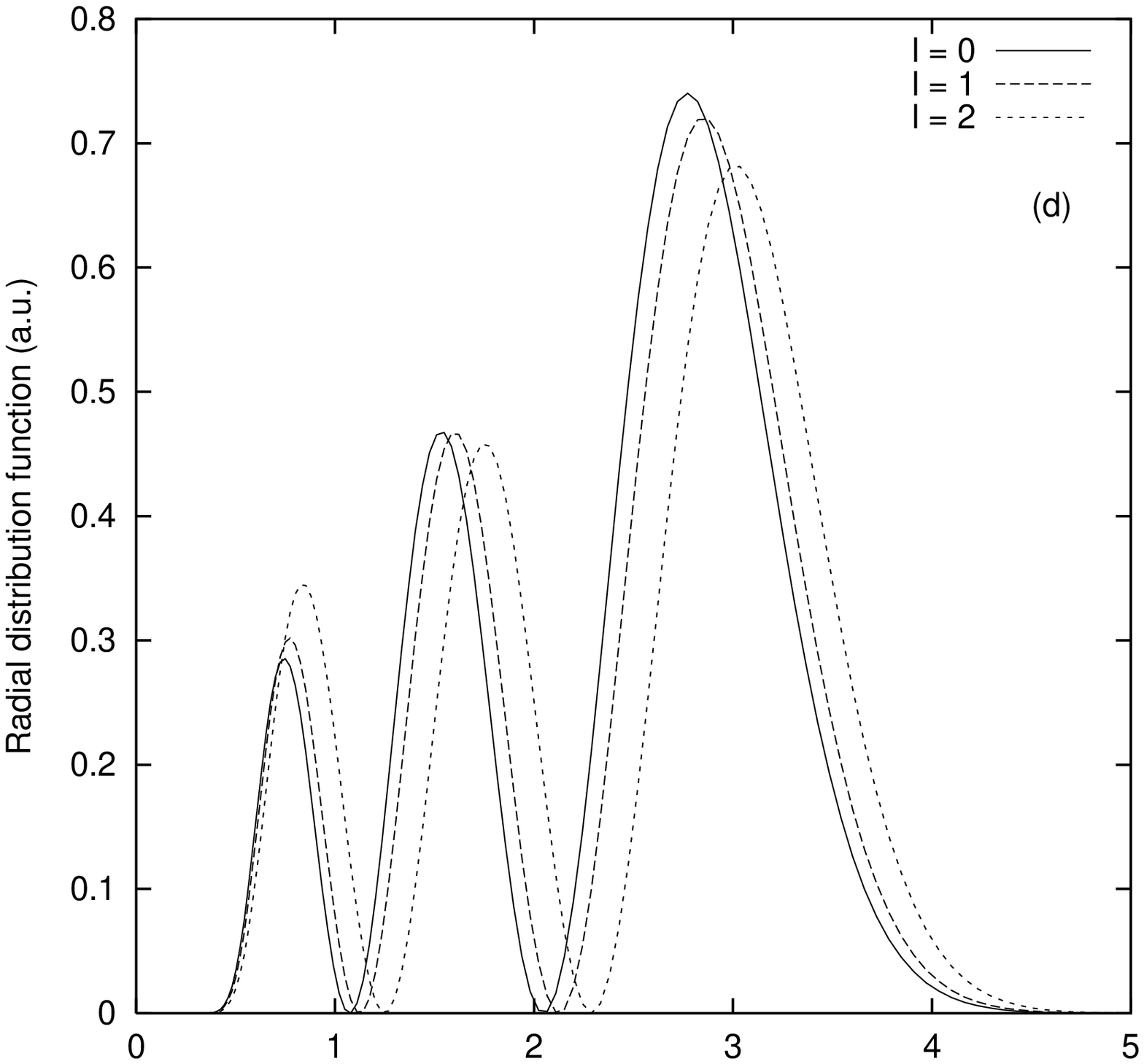}
\end{minipage}
\hspace{0.15in}
\begin{minipage}[b]{0.40\textwidth}\centering
\includegraphics[scale=0.25]{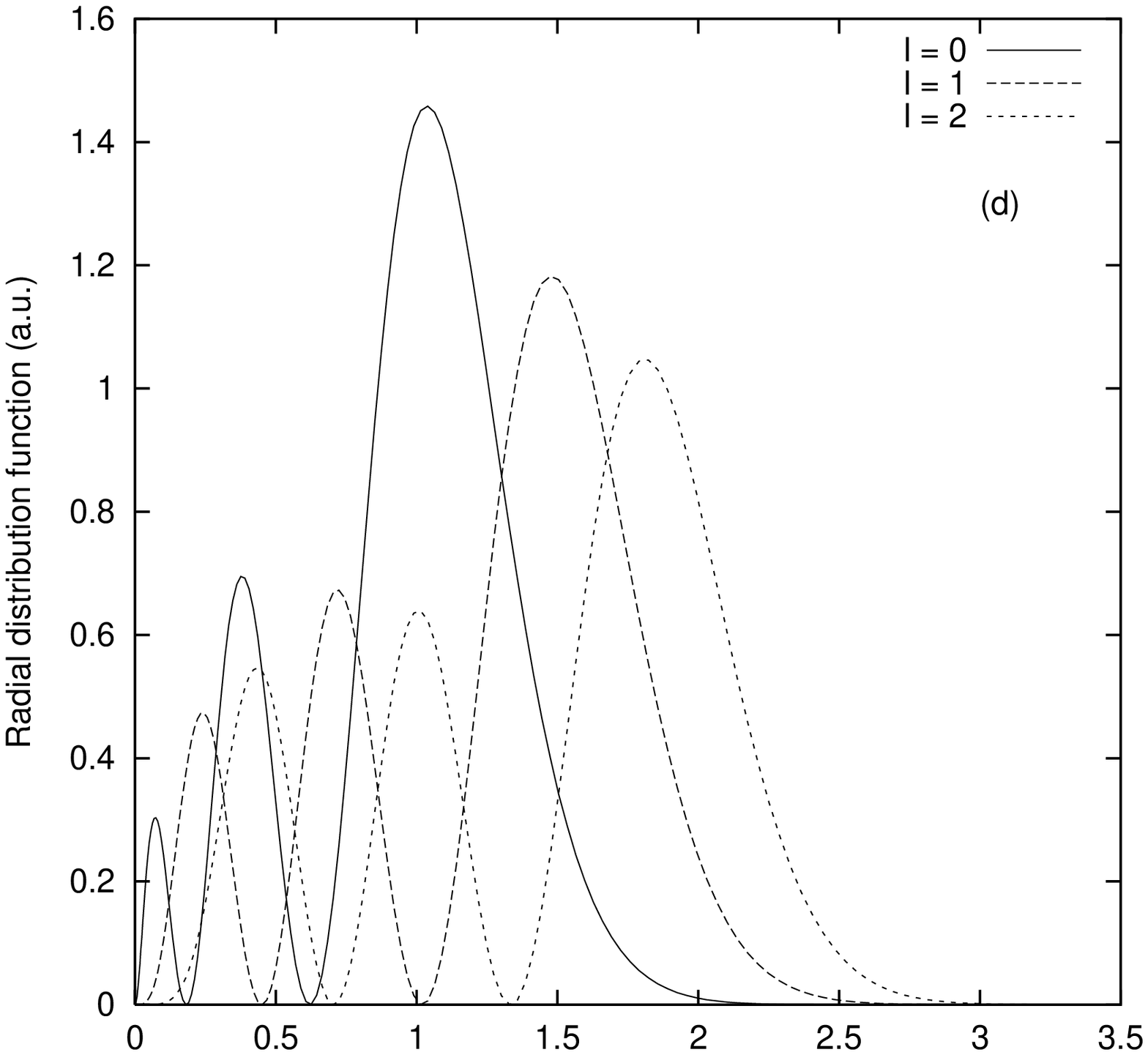}
\end{minipage}
\\
\begin{minipage}[b]{0.40\textwidth}\centering
\includegraphics[scale=0.25]{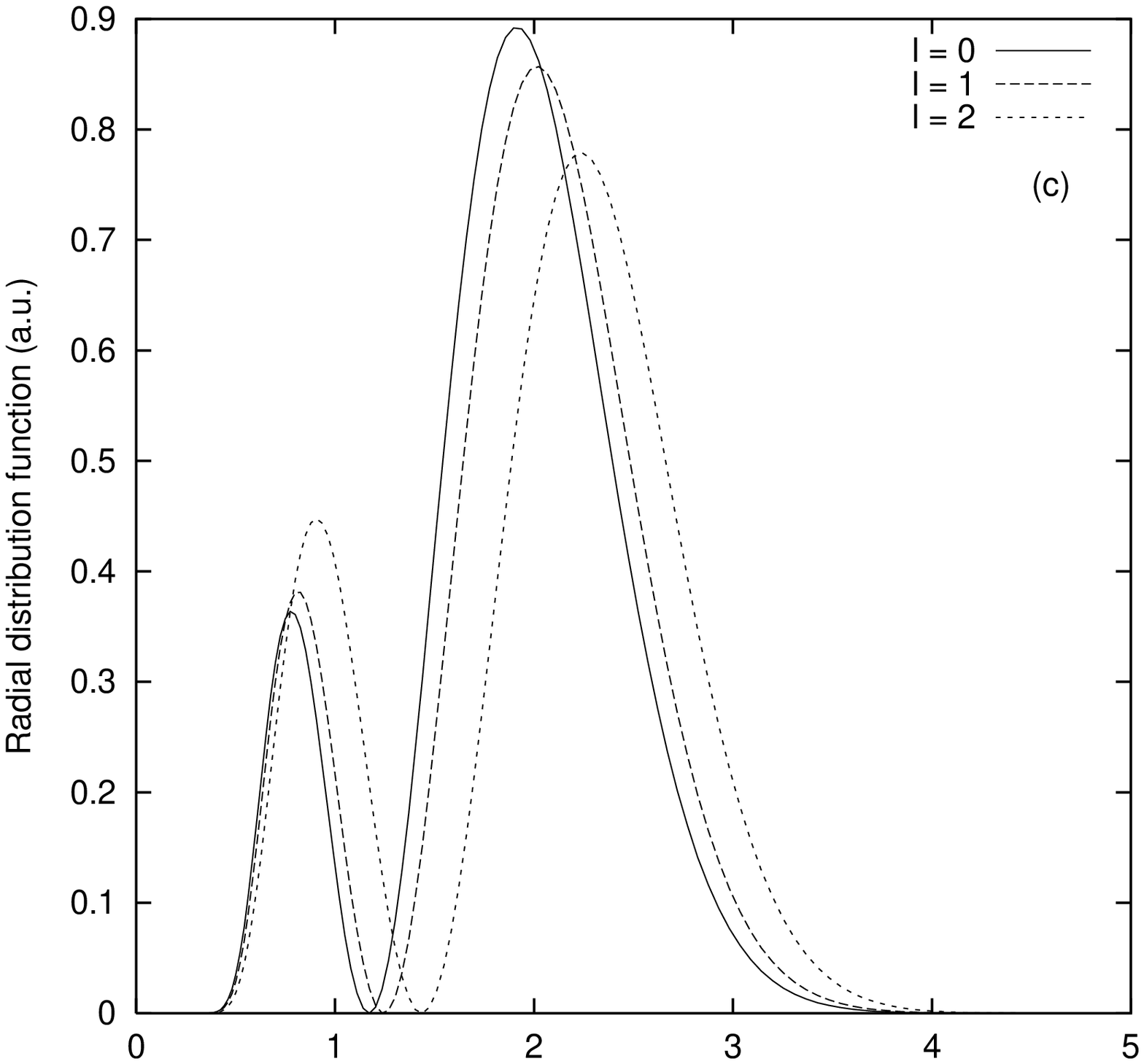}
\end{minipage}
\hspace{0.15in}
\begin{minipage}[b]{0.40\textwidth}\centering
\includegraphics[scale=0.25]{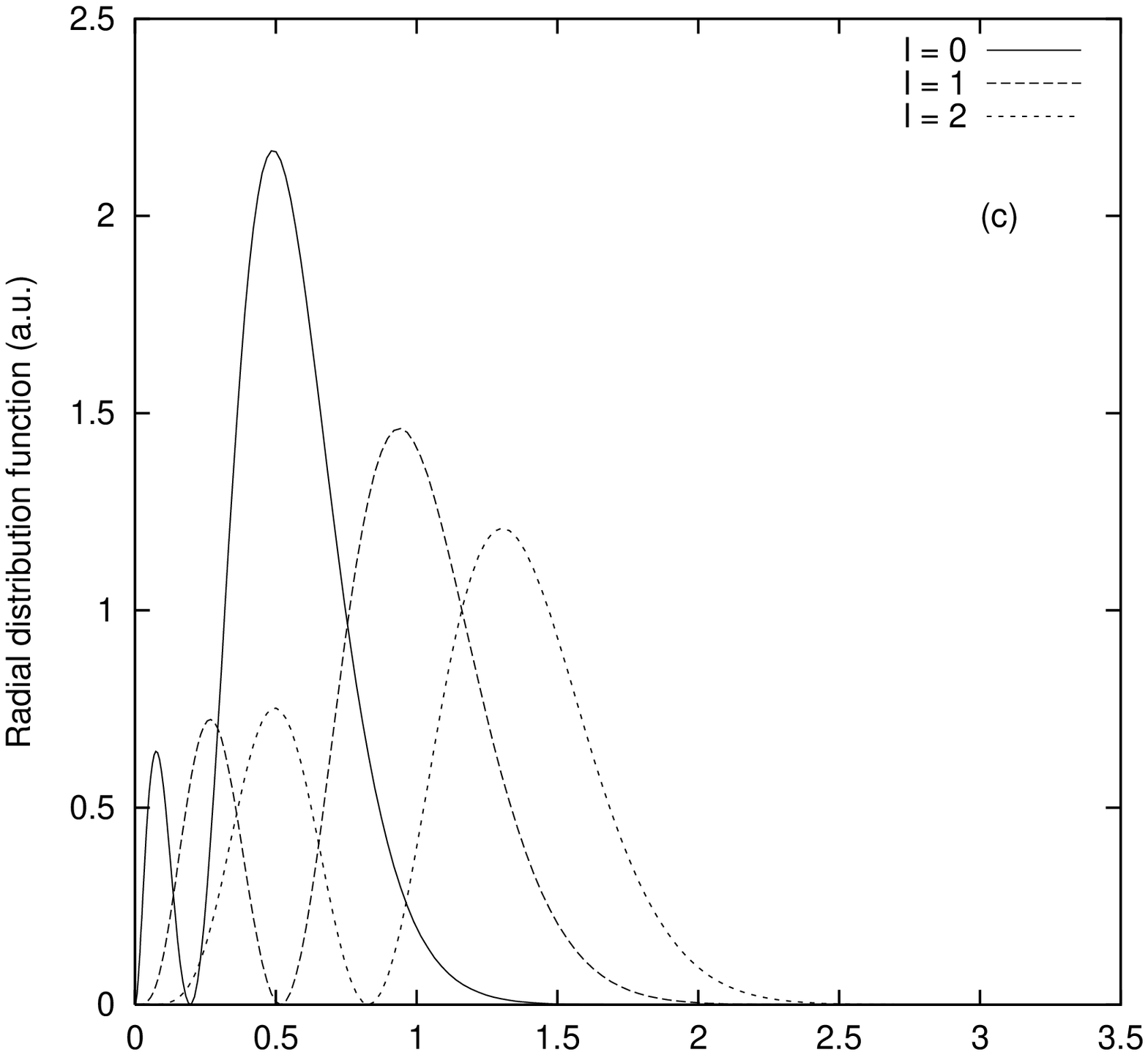}
\end{minipage}
\\
\begin{minipage}[b]{0.40\textwidth}\centering
\includegraphics[scale=0.25]{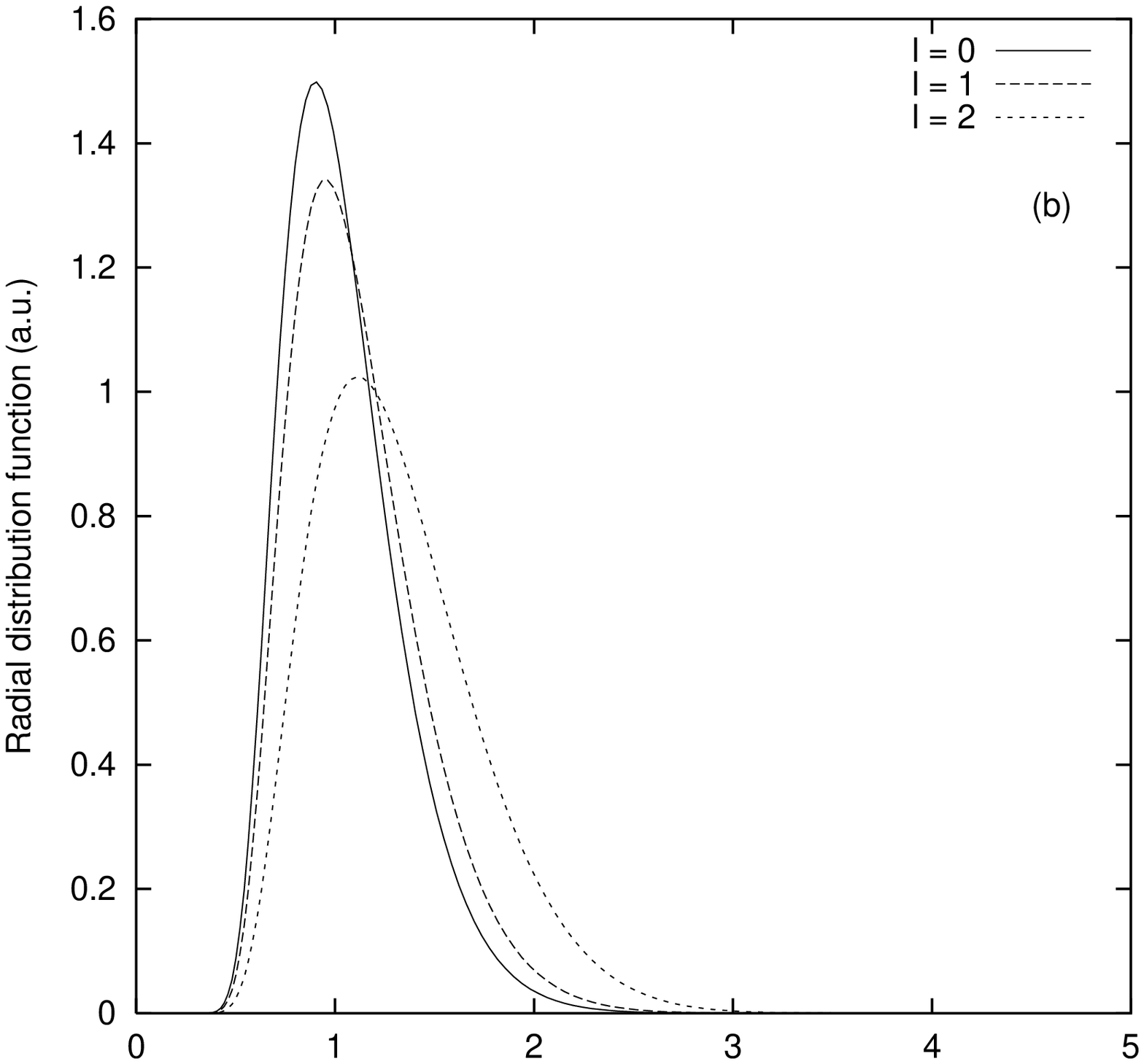}
\end{minipage}
\hspace{0.15in}
\begin{minipage}[b]{0.40\textwidth}\centering
\includegraphics[scale=0.25]{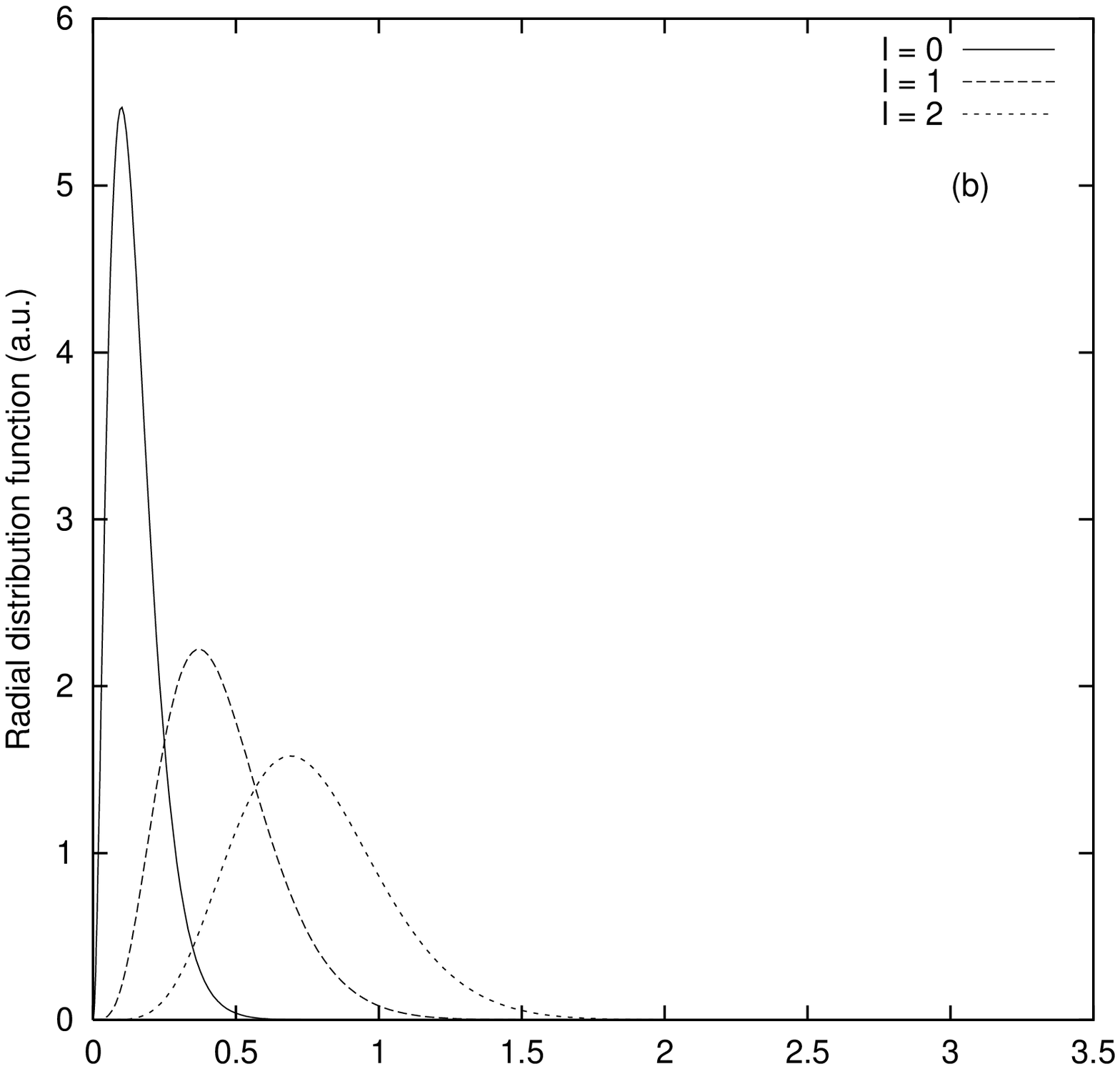}
\end{minipage}
\\
\begin{minipage}[b]{0.40\textwidth}\centering
\includegraphics[scale=0.25]{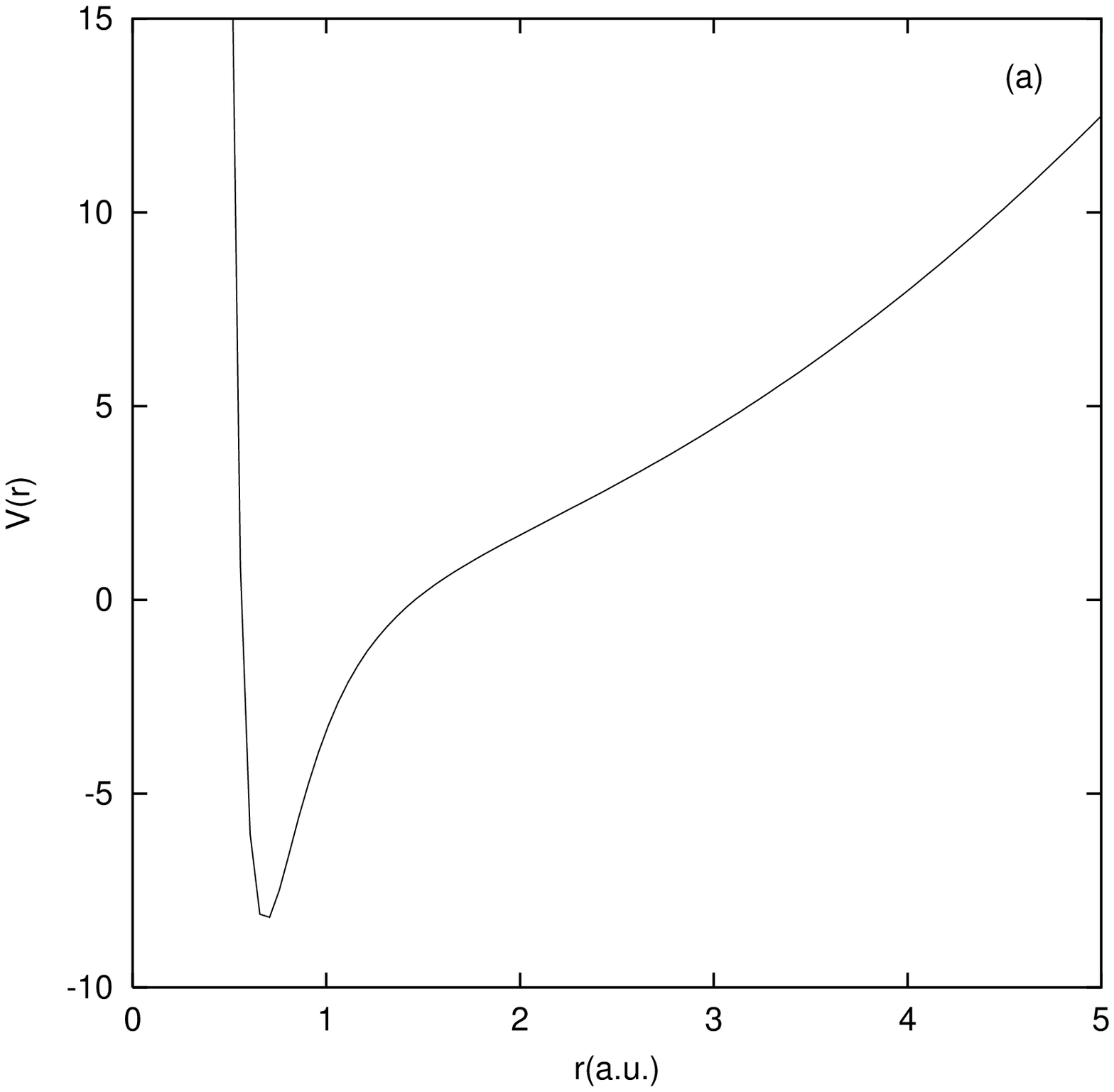}
\end{minipage}
\hspace{0.15in}
\begin{minipage}[b]{0.40\textwidth}\centering
\includegraphics[scale=0.25]{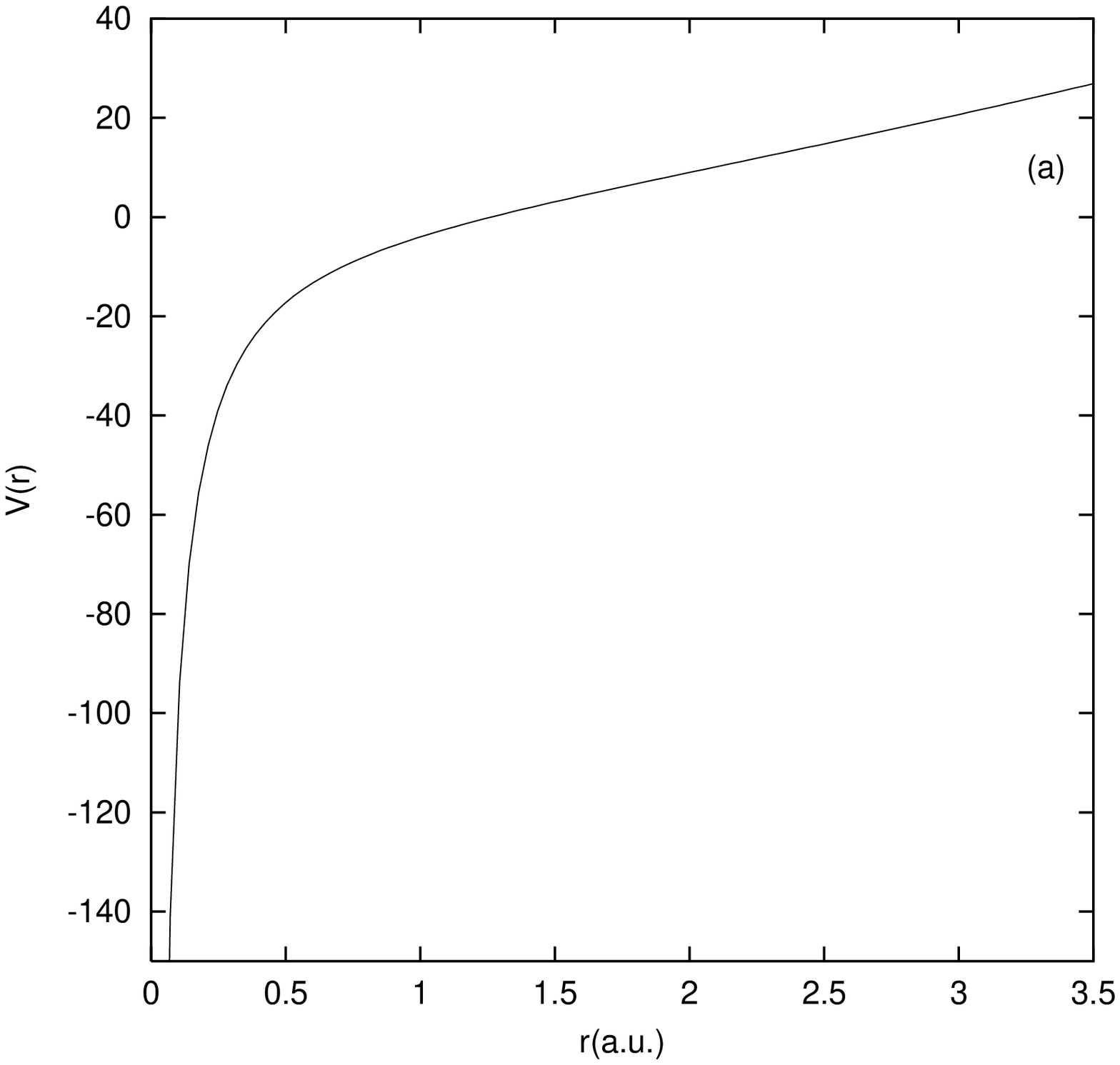}
\end{minipage}

\caption{The radial probability distribution functions, $|rR_{n\ell}|^2$ for the potentials 
in Eqs.~ (1) and (2) in left and right panel respectively. The first four states corresponding
to $\ell=0,1,2$ are shown: (a) the potential, (b) the ground, (c) first excited, and (d) the 
second excited state. The parameters are $a=0.5, b=-5.625, c=1.7578125.$ and 
$Z=10, g=5, \lambda=1.$}\label{fig:fig}
\end{figure}

Before passing, a few remarks should be made. In the present method, no {\em unphysical} states 
are obtained. It is worthwhile to note that some of the methods employed for these kind of 
potentials in the literature often have the unwanted feature of producing unphysical roots, 
e.g., the Ricatti-Pad\'e scheme for a class of singular potentials, commonly termed as spiked 
oscillators. On the other hand, existence of false (unphysical) eigenvalues in the 
Hill-determinant method for a perturbed oscillator or perturbed Coulomb potential has been 
suggested by some authors on mathematical grounds. Later it was pointed out [43] that these 
unphysical or false states manifest themselves by having the negative values for 
$\langle r \rangle$ or $\langle r^{-1} \rangle$. Some of the well-known methods have another 
undesirable feature that they give poor or erroneous results for potentials having multiple 
wells, for certain parameter sets or sometimes even for certain states within a particular 
parameter set. As an example, the shifted 1/N method for the perturbed Coulomb potential [22] 
fails to give any result when any four of the five quantities $Z$, $\lambda$, $g$, $\ell,$ 
$n_r$ are kept fixed and only the fifth one is varied. The present method is {\em general} in 
the sense that it is applicable to a wide spectrum of parameters without requiring any special 
relation among them, irrespective of the strength of the coupling involved and therefore lifts 
any such restrictions. In view of the simplicity and accuracy offered by this method for the 
polynomial as well as the singular potentials considered in this work for both ground and higher 
excited states, it may be hoped that this prescription may be useful for a wide range of other 
similar potentials of interest. Thus other singular potentials having higher order terms in 
Eqs. (1) or (2) or the nonsingular polynomial potentials with higher order anharmonicities like 
quartic, sextic, octic, etc., may as well be treated by this method very well. Finally, we 
mention that while practically all of the common methodologies in quantum mechanics for such 
systems involve the solution of the time-independent Schr\"odinger equation, recently a 
time-dependent (TD) formalism [44-47] has been proposed as well for the accurate calculation of 
static properties, which is, in principle, exact. Applications have been made for the ground 
and excited states of double well and anharmonic oscillators having higher order terms, 
multiple-well and self-interacting oscillators in one dimension [44-46], as well as the 
ground-state electronic properties of noble gas atoms [47] through an amalgamation of the 
density functional and quantum fluid dynamical treatment. It may be worthwhile to study the 
performance of such TD methods for the singular potentials dealt in this work.   

\section{Conclusion}
An {\em accurate} and {\em simple} methodology has been proposed for certain singular 
potentials employing the generalized pseudospectral method. The prescription is {\em general} 
and {\em reliable}. The energies, densities and expectation values presented for ground and 
excited states of the (i) harmonic potential with inverse quartic and sextic perturbation and 
(ii) the perturbed Coulomb potential with linear and quadratic terms are reported for a broad 
spectrum of potential parameters. This scheme is capable of producing very good quality results 
for both ground and higher excited states. The $v=48,49,\cdots ,48,49\ $ states corresponding 
to the angular momentum quantum numbers $\ell=0,1,\cdots ,8,9\ $ for the pure three dimensional 
quartic oscillator illustrates this fact. Many states have been reported here for the first 
time which might constitute a useful set for future referencing. A large number of states are 
shown to be in excellent agreement with those from other sophisticated methods available in 
the literature and augurs well for its practical applicability to other similar potentials of 
physical and mathematical interest in quantum mechanics. 

\begin{acknowledgments}
I gratefully acknowledge the hospitality provided by the University of New Brunswick, 
Fredericton, NB, Canada. 
\end{acknowledgments}


\begin{thebibliography}{99}
\bibitem{1} J.\ R.\ Klauder, Phys.\ Lett.\ B \textbf{47B} 523 (1973).
\bibitem{2} L.\ C.\ Detwiler and J.\ R.\ Klauder, Phys.\ Rev.\ D \textbf{11} 1436 (1975).
\bibitem{3} M.\ Znojil, J.\ Math.\ Phys.\ \textbf{30} 23 (1989).
\bibitem{4} E.\ M.\ Harrell, Ann.\ Phys.,\ NY \textbf{105} 379 (1977).
\bibitem{5} M.\ Znojil, J.\ Phys.\ A \textbf{25} 2111 (1982).
\bibitem{6} M.\ Znojil, J.\ Math.\ Phys.\ \textbf{31} 108 (1990).
\bibitem{7} R.\ Guardiola and J.\ Ros, J.\ Phys.\ A \textbf{25} 1351 (1992).
\bibitem{8} V.\ C.\ Aguilera-Navarro, G.\ A.\ Est\'evez and R.\ Guardiola, J.\ Math.\ 
Phys.\ \textbf{31} 99 (1990).
\bibitem{9} E.\ Buendi\'a, F.\ J.\ G\'alvez and A.\ Puertas, J.\ Phys.\ A \textbf{28} 
6731 (1995). 
\bibitem{10} J.\ Killingbeck, J.\ Phys.\ B \textbf{15} 829 (1982).
\bibitem{11} J.\ P.\ Killingbeck, G.\ Jolicard, and A.\ Grosjean, J.\ Phys.\ A 
\textbf{34} L367 (2001).
\bibitem{12} M.\ Landtman, Phys.\ Lett.\ A \textbf{175} 147 (1993).
\bibitem{13} R.\ S.\ Kaushal, Ann.\ Phys.,\ NY \textbf{206} 90 (1991).
\bibitem{14} R.\ S.\ Kaushal and D.\ Parashar, Phys.\ Lett.\ A \textbf{170} 335 (1992).
\bibitem{15} B.\ Chakrabarti and T.\ K.\ Das, J.\ Phys.\ A \textbf{35} 4701 (2002). 
\bibitem{16} E.\ R.\ Vrscay, Phys.\ Rev.\ A \textbf{31} 2054 (1985).
\bibitem{17} C.\ Quigg and J.\ L.\ Rosner, Phys. Rep.\ \textbf{56} 167 (1979).
\bibitem{18} E.\ Eichten, E.\ Gottfried, T.\ Kinoshita, K.\ D.\ Lane and T.\ M.\ Yan, 
Phys.\ Rev.\ D \textbf{17} 3090 (1978).
\bibitem{19} J.\ E.\ Avron, Ann.\ Phys.\ \textbf{131} 73 (1991). 
\bibitem{20} S.\ Skupsky, Phys.\ Rev.\ A \textbf{21} 1316 (1980).
\bibitem{21} D.\ Bessis, E.\ R.\ Vrscay and C.\ R.\ Handy, J.\ Phys.\ A \textbf{20}, 
419 (1987).
\bibitem{22} R.\ K.\ Roychoudhury and Y.\ P.\ Varshni, J.\ Phys.\ A \textbf{21} 3025 (1988).
\bibitem{23} R.\ P.\ Saxena and V.\ S.\ Varma, J.\ Phys.\ A \textbf{15} L149 (1982); 
\emph {ibid.} L221 (1982).
\bibitem{24} D.\ P.\ Datta and S.\ Mukherjee, J.\ Phys.\ A \textbf{13} 3161 (1980).
\bibitem{25} M.\ Znojil, J.\ Phys.\ A \textbf{16} 213 (1983).
\bibitem{26} R.\ N.\ Chaudhuri and M.\ Mondal, Phys.\ Rev.\ A \textbf{52} 1850 (1995).
\bibitem{27} R.\ Adhikari, R.\ Dutt and Y.\ P.\ Varshni, Phys.\ Lett.\ A \textbf{141} 
1 (1989).
\bibitem{28} E.\ Castro and P.\ Mart\'in, J.\ Phys.\ A \textbf{33} 5321 (2000).
\bibitem{29} A.\ K.\ Roy and S.\ I.\ Chu, Phys.\ Rev.\ A \textbf{65} 052508 (2002).
\bibitem{30} A.\ K.\ Roy, J.\ Phys.\ B \textbf{37} 4369 (2004).
\bibitem{31} See, for example, R.\ G.\ Parr and W.\ Yang, {\em Density-functional theory of 
atoms and molecules} (Oxford University Press, NY) (1989).
\bibitem{32} G.\ Yao and S.\ I.\ Chu, Chem.\ Phys.\ Lett.\ \textbf{204} 381 (1993).
\bibitem{33} J.\ Wang, S.\ I.\ Chu and C.\ Laughlin, Phys.\ Rev.\ A \textbf{50} 3208 (1994).
\bibitem{34} X.\ M.\ Tong and S.\ I.\ Chu, Phys.\ Rev.\ A \textbf{64} 013417 (2001).
\bibitem{35} A.\ K.\ Roy and S.\ I.\ Chu, Phys.\ Rev.\ A \textbf{65} 043402 (2002).
\bibitem{36} A.\ K.\ Roy, J.\ Phys.\ G \textbf{30} 269 (2004).
\bibitem{37} A.\ K.\ Roy, Phys.\ Lett. A \textbf{321} 231 (2004).
\bibitem{38} A.\ K.\ Roy, Pramana-J.\ Phys.\ (in press).
\bibitem{39} S.\ Bell, R.\ Davidson and P.\ A.\ Warsop, J.\ Phys.\ B \textbf{3} 123 (1970).
\bibitem{40} P.\ M.\ Mathews, M.\ Seetharaman and S.\ Raghavan, J.\ Phys.\ A \textbf{15} 
103 (1982).
\bibitem{41} A.\ Holubec, A.\ D.\ Stauffer, P.\ Acacia and J.\ A.\ Stauffer, J. Phys.\ A 
\textbf{23} 4081 (1990). 
\bibitem{42} Y.\ P.\ Varshni, Phys.\ Lett.\ A \textbf{183} 9 (1993).
\bibitem{43} J.\ Killingbeck, J.\ Phys.\ A \textbf{19} 2903 (1986).
\bibitem{44} A.\ K.\ Roy, N.\ Gupta and B.\ M.\ Deb, Phys.\ Rev.\ A \textbf{65} 012109 
(2002).
\bibitem{45} N.\ Gupta, A.\ K.\ Roy and B.\ M.\ Deb, Pramana-J.\ Phys.\ \textbf{59} 575 
(2002).
\bibitem{46} A.\ Wadehra, A.\ K.\ Roy and B.\ M.\ Deb, Int.\ J.\ Quant.\ Chem.\ \textbf{91} 
597 (2003). 
\bibitem{47} A.\ K.\ Roy and S.\ I.\ Chu, J.\ Phys.\ B \textbf{35} 2075 (2002).
\end{thebibliography}
\end{document}